\documentclass[conference]{IEEEtran}
\IEEEoverridecommandlockouts

\usepackage{cite}
\usepackage{amsmath,amssymb,amsfonts}
\usepackage{algorithmic}
\usepackage{graphicx}
\usepackage{textcomp}
\usepackage{xcolor}
\usepackage{booktabs}
\usepackage{easyReview}
\def\BibTeX{{\rm B\kern-.05em{\sc i\kern-.025em b}\kern-.08em
    T\kern-.1667em\lower.7ex\hbox{E}\kern-.125emX}}

\begin{document}

\title{LLM-enhanced Air Quality Monitoring Interface via Model Context Protocol\\
}

\author{\IEEEauthorblockN{1\textsuperscript{st} Yu-Erh Pan}
\IEEEauthorblockA{
\textit{University of Wisconsin-Milwaukee}\\
Milwaukee, WI, U.S. \\
yuerhpan@uwm.edu}
\and
\IEEEauthorblockN{2\textsuperscript{nd} Ayesha Siddika Nipu}
\IEEEauthorblockA{
\textit{University of Wisconsin-Milwaukee}\\
Milwaukee, WI, U.S. \\
nipua@uwm.edu}
}

\maketitle

\begin{abstract}
Air quality monitoring is central to environmental sustainability and public health, yet traditional systems remain difficult for non-expert users to interpret due to complex visualizations, limited interactivity, and high deployment costs. Recent advances in Large Language Models (LLMs) offer new opportunities to make sensor data more accessible, but their tendency to produce hallucinations limits reliability in safety-critical domains. To address these challenges, we present an LLM-enhanced Air Monitoring Interface (AMI) that integrates real-time sensor data with a conversational interface via the Model Context Protocol (MCP). Our system grounds LLM outputs in live environmental data, enabling accurate, context-aware responses while reducing hallucination risk. The architecture combines a Django-based backend, a responsive user dashboard, and a secure MCP server that exposes system functions as discoverable tools, allowing the LLM to act as an active operator rather than a passive responder. Expert evaluation demonstrated high factual accuracy (4.78), completeness (4.82), and minimal hallucinations (4.84), on a scale of 5, supported by inter-rater reliability analysis. These results highlight the potential of combining LLMs with standardized tool protocols to create reliable, secure, and user-friendly interfaces for real-time environmental monitoring.
\end{abstract}


\section{Introduction}

With the growing emphasis on healthy living, air quality monitoring has become a significant area of research. Numerous studies have shown that indoor air pollutants can significantly affect human health \cite{pillarisetti2022indoor}. Particulate Matter (PM) suspended in the air, particularly PM2.5, has been proven to be directly associated with various health risks including cardiovascular diseases, respiratory infections, and even cancer \cite{world2021global}. Therefore, developing air quality monitoring systems that are real-time, accurate, and easy to understand is crucial for enhancing public health awareness and formulating effective environmental policies.

Existing air quality monitoring systems still face numerous challenges in practical applications. Traditional monitoring stations, while capable of providing high-precision data, have high deployment costs and limited coverage \cite{colleaux2025air}, resulting in insufficient spatial and temporal resolution of data \cite{cromar2019air}. More importantly, these systems primarily present data through complex charts and indicators, making it difficult for ordinary users without professional backgrounds to intuitively understand the meaning behind the data and make informed decisions accordingly. Furthermore, traditional systems generally have poor interactivity, preventing users from querying specific questions of interest in a natural and direct manner, thereby limiting the practical value of data and public engagement \cite{saini2020comprehensive}.

In recent years, artificial intelligence technologies represented by Large Language Models (LLMs) have provided new opportunities to overcome the aforementioned challenges. The excellent natural language processing and generation capabilities of LLMs enable them to transform complex raw data into easily understandable textual summaries, analyses, and recommendations, significantly lowering the barrier for non-professional users to access information. Through natural language interaction interfaces, users can directly ask questions and receive real-time, context-aware responses. The reasoning capabilities of LLMs can also be used for advanced analytical tasks such as anomaly detection and trend prediction, thereby endowing traditional monitoring systems with unprecedented levels of intelligence.

\begin{figure}[tb]
    \centering
    \includegraphics[width=\columnwidth]{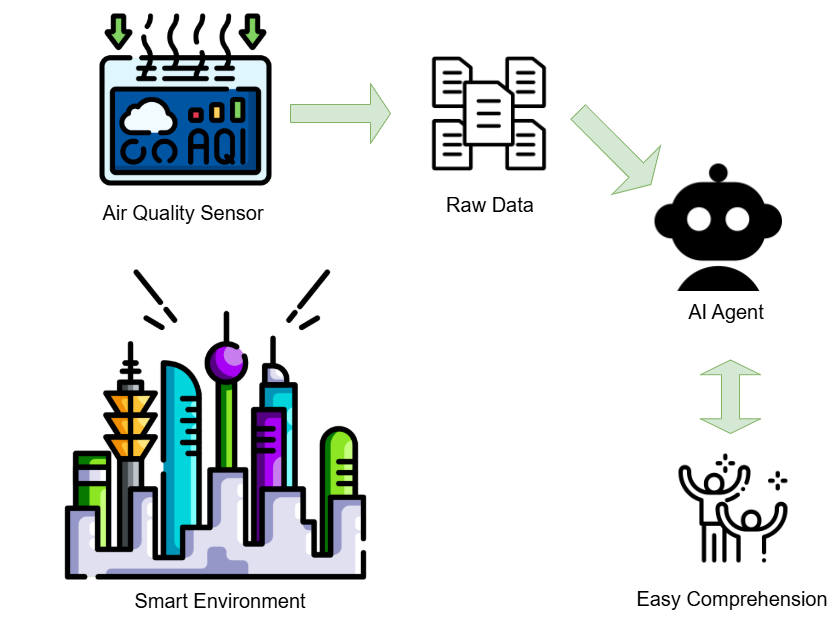}
    \caption{A smart air quality monitoring system}
    \label{fig:smart-environment}
\end{figure}

Despite the tremendous potential of LLMs in the field of data analysis, reliably integrating them with real-time environmental monitoring data remains an under-explored research topic. Traditional LLM applications often suffer from ``hallucinations" due to lack of real-time contextual knowledge, providing inaccurate or outdated information \cite{huang2025survey}. Unlike traditional retrieval-based methods, our system leverages the Model Context Protocol (MCP) to allow LLMs to securely access real-time IoT data, ensuring grounded, context-aware responses \cite{hou2025model}. Figure~\ref{fig:smart-environment} illustrates the overall architecture of our proposed system. Our main contributions can be summarized as follows:
\begin{itemize}
\item \textbf{System Design and Implementation}: We propose and develop a complete, web-based air quality monitoring system for receiving, storing, and visualizing diverse sensor data in real time.
\item \textbf{Innovative Integration of LLM and MCP}: A novel approach for enabling the system to interact with real-time data through an LLM connected via MCP, supporting intelligent query handling and data-driven insights.
\end{itemize}

\section{Related Work}

\textbf{Air Quality Monitoring and Accessibility:}  
Traditional air quality monitoring systems have matured in terms of sensor networks and data acquisition. Saini et al. provided a comprehensive overview of indoor air quality monitoring systems, documenting advances in sensor network design, pollutant detection, and data collection mechanisms \cite{saini2020comprehensive}. These systems often achieve impressive technical performance. However, Gupta et al. shown that the information presented by many digital platforms is not always interpretable to non-expert audiences, revealing a significant gap between data acquisition and user-friendly communication of results \cite{gupta2022human}. Morawska et al. emphasized the rapid proliferation of low-cost sensors and community-driven monitoring networks, which greatly improve spatial and temporal resolution, yet, the sheer volume and heterogeneity of data can overwhelm typical interfaces, reducing their effectiveness for general users \cite{morawska2018applications}. Recently, Aguado et al. evaluated low-cost indoor air quality sensors in real-world settings for CO$_2$ and particulate matter, and while the devices supported trend tracking, the study underscores the need for intuitive feedback systems that help non-experts understand what the readings mean for health and behavior \cite{aguado2025verification}. Taken together, this body of work shows that although sensor technology has matured and become more affordable, the critical challenge remains: designing interfaces that can translate dense, technical measurements into actionable insights that non-technical users can easily grasp and act upon.

\textbf{LLMs for Conversational Interfaces:} 
Recent advances in LLMs have introduced new opportunities to make complex data more accessible through conversational interfaces. By allowing natural language queries, LLMs lower the barrier for non-technical users. However, In two separate studies, Dam et al. and Huang et al. both emphasized a central challenge: the propensity of LLMs for hallucination, where models produce fluent but factually incorrect responses when operating without sufficient context \cite{dam2024complete,huang2025survey}. To mitigate this, Brown et al. demonstrated that directly injecting sensor data into the model's prompt could enable context-aware summaries and answers about specific datasets \cite{brown2020language}. While effective at small scale, their approach is constrained by the finite context windows of current LLMs. More recently, Islam et al. proposed a prompt ensemble strategy that improved reliability of LLMs response by aggregating outputs from multiple prompts, offering a promising pathway to reduce hallucination and enhance robustness in domain-specific applications \cite{islam2025llm}.

\textbf{Retrieval Augmented Generation for Scalability:}
The scalability limitations of direct context injection have been well established, most notably through the “lost-in-the-middle” phenomenon, where model performance degrades when long sequences are provided \cite{liu2023lost}. To address this issue, Gao et al. introduced Retrieval-Augmented Generation (RAG) as a dominant framework \cite{gao2023retrieval}. RAG operates by retrieving a targeted subset of relevant information from a larger knowledge base and passing only this filtered context to the LLM. While this approach has proven effective across a wide range of applications, Barnett et al. pointed out that its success is highly sensitive to the quality of the retrieval process, including choices around chunking strategy and parameter tuning \cite{barnett2024seven}. Furthermore, RAG is not ideally suited for structured, time-series data such as environmental sensor streams, where semantic chunking fails to capture the inherent temporal dependencies in the data.

\textbf{Toward Agentic Systems with MCP:}  
Building on the limitations of retrieval-based methods, our work adopts an active, agentic framework. Rather than having the system predefine which information to supply, the decision is delegated directly to the model. To enable this, Anthropic et al. introduced the Model Context Protocol (MCP), which allows LLMs to autonomously call a predefined set of secure tools to interact with backend systems \cite{anthropic2024mcp}. This approach provides two key advantages over RAG: first, it ensures data fidelity and completeness by granting models direct access to raw, real-time data sources; second, it shifts the model’s role from a passive question-answerer to an active system operator. To the best of our knowledge, applying a standardized protocol such as MCP to ground LLMs in real-time IoT monitoring remains a novel and underexplored research area, which our work aims to address.

\section{Methodology}

This section presents the system design and technical implementation of our proposed LLM-enhanced Air Monitoring Interface (AMI). The platform integrates IoT sensor data with a conversational interface powered by LLM, enabling non-expert users to interact with air quality information in natural language. We used MCP to ground LLM responses in real-time sensor data, thereby reducing hallucination and extending the model's functionality beyond static Q\&A. The overall design aims to establish an efficient, scalable, and highly secure intelligent interactive system that can seamlessly integrate LLM to empower traditional IoT monitoring applications.

\subsection{System Architecture}
\begin{figure}[tb]
    \centering
    \includegraphics[ width= 0.88 \columnwidth]{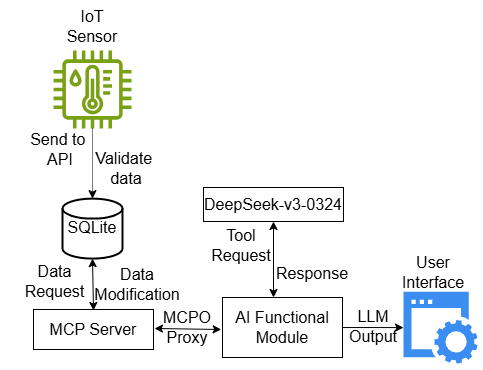}
    \caption{System architecture of the AMI platform, illustrating interactions from sensor data acquisition to the LLM-driven interface.}
    \label{fig:system_architecture}
\end{figure}

The AMI adopts a multi-layered architecture built on the Django web framework, ensuring modularity and separation of concerns as reflected in Figure~\ref{fig:system_architecture}. The major components are as follows:

\begin{itemize}
    \item \textbf{Data Acquisition Layer:} Collects real-time air quality readings via HTTP POST requests from IoT sensors to a RESTful API endpoint (\texttt{/sensor\_data/}). The data include temperature, humidity, CO$_2$, PM$_{1.0}$, PM$_{2.5}$, and PM$_{10}$, which are validated before storage.
    \item \textbf{Backend Service:} Implements business logic, user authentication, and data management in python-based Django framework. It also acts as a bridge between the frontend interface and the AI module.
    \item \textbf{Database:} A lightweight SQLite database persists sensor histories, user profiles, and configuration data, supporting rapid prototyping and portability.
    \item \textbf{Frontend Interface:} A responsive dashboard built with HTML, CSS, and JavaScript enables real-time visualization, historical trend analysis, and agentic-AI interaction.
    \item \textbf{AI Functional Module:} Uses the \texttt{openai} Python library to communicate with the DeepSeek-V3-0324 model, orchestrating dialogue flow and MCP tool calls.
    \item \textbf{MCP Server:} A standalone Python service based on FastMCP. It exposes backend functions (e.g., data queries, issue reporting, profile edits) as standardized MCP tools, automatically converted to OpenAPI format with the \texttt{mcpo} library for discovery by the LLM.
\end{itemize}

\subsection{Core Interaction Protocol: MCP}
The Model Context Protocol (MCP)\footnote{https://modelcontextprotocol.io/docs/learn/architecture} is a client–server framework that standardizes how llms interact with external tools and data sources. It defines a JSON-RPC based data exchange layer and supports multiple transport mechanisms, such as standard input/output for local servers or HTTP for remote ones. Within this architecture, the AI application (host) communicates with one or more MCP servers through dedicated MCP clients, enabling structured discovery of available tools, resources, and prompts. This separation of responsibilities ensures that the model can request context or invoke actions dynamically, without embedding tool logic into the model itself.

MCP serves as the integration backbone of our system, enabling secure and standardized communication between the LLM and backend tools. Unlike traditional ad hoc integrations, MCP provides a unified mechanism that improves interoperability, maintainability, and extensibility. Backend functions such as data queries and issue reporting are exposed as discoverable MCP tools, which lowers integration complexity and decouples AI logic from backend code. This design allows future replacement or addition of LLMs without rebuilding tool interfaces. Moreover, by grounding model outputs in real-time sensor operations, MCP transforms the LLM into an active operator and significantly reduces hallucinations caused by static or outdated knowledge. The detailed MCP interaction flow is illustrated in Figure~\ref{fig:mcp_interaction}. The interaction flow begins with a user request. The backend retrieves current tool definitions from the MCP agent, forwards them with the request to the LLM, and executes any tool calls selected by the model. Results are returned as context, enabling the model to generate precise and grounded responses.

\subsection{LLM Integration and Prompt Engineering}
DeepSeek-V3-0324 was chosen for its strong tool-use abilities, cost efficiency, and compatibility with the OpenAI API, which accelerated integration. To ensure reliability and security, a multi-layered prompt and code-level enforcement strategy was designed:

\begin{figure}[tb]
    \centering
    \includegraphics[width= 0.88 \columnwidth]{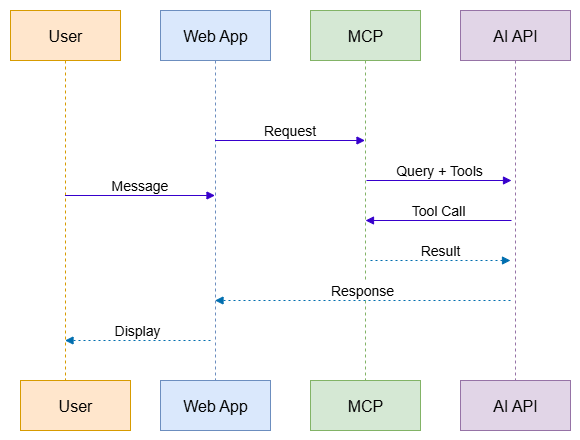}
    \caption{Sequence diagram of the MCP interaction flow, showing how user requests translate into LLM tool calls executed via the MCP server.}
    \label{fig:mcp_interaction}
\end{figure}

\subsubsection{System Prompt}
The system prompt is a set of high-level instructions embedded in the backend application, designed to define the LLM's core role and behavioral boundaries. It instructs the model to act as an intelligent assistant for the AMI system, with the primary responsibility of assisting the currently logged-in user and strictly adhering to permission constraints when calling tools, prohibiting access to or modification of other users' information. This prompt ensures that the model consistently maintains its professionalism and functionality during interactions.

\subsubsection{MCP Tool Prompt}
Through the \texttt{tools} parameter of the \texttt{openai} library, we provide the LLM with precise tool definitions. These definitions, converted from OpenAPI format, include each tool's name, functional description, and JSON Schema-based parameters. This precise tool prompting enables the LLM to clearly understand when and how to correctly invoke backend functions.

\subsubsection{Defense-in-Depth: A Dual Security Mechanism of Prompt and Code}
Relying solely on prompt guidance is insufficient to address all security risks. Therefore, we implemented a defense-in-depth strategy that enforces user authentication at the code level. For all MCP tools involving sensitive operations, the system backend forcibly overwrites the user identity parameters in tool calls, ensuring they always match the currently authenticated logged-in user. This mechanism fundamentally eliminates risks of impersonation and unauthorized data access, ensuring that even if the LLM is misled or maliciously guided, it cannot bypass the system's permission controls. However, we also recognize that this highlights the potential risks of completely trusting tool calls initiated by LLMs when handling sensitive operations, necessitating the design of such backend security mechanisms for reinforcement.

\begin{figure}[tb]
    \centering
    \includegraphics[width=0.8 \columnwidth]{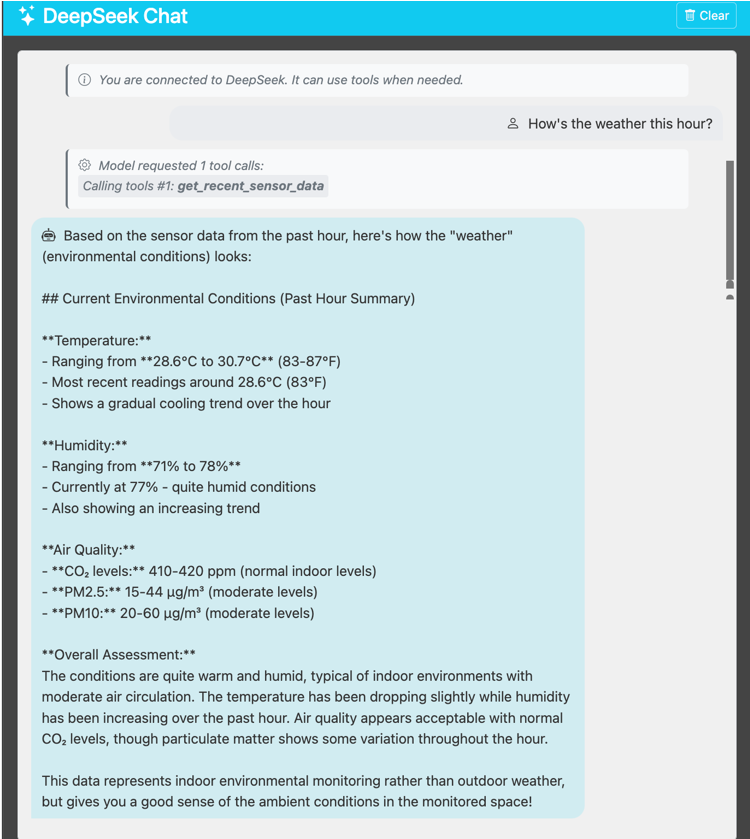}
    \caption{An example of a real-time data query. The user asks a question in natural language, and the LLM invokes the appropriate tool to fetch and summarize live sensor data.}
    \label{fig:data_query}
\end{figure}

\section{Results and Discussion}
This section presents the results of our user evaluation, which was designed to assess the quality and reliability of the LLM-driven responses from our AMI platform. We first present the quantitative results from our expert evaluation and then discuss the interpretation and implications of these findings.

\subsection{System Interface}

This subsection visually demonstrates the core functionalities of the AMI platform and the user interaction flow through a series of interface screenshots. These examples highlight how the integration of an LLM via MCP enables intuitive, powerful, and secure interactions.

Figure~\ref{fig:data_query} illustrates a typical data query scenario, which is the system's primary use case. When a user asks a general question about the current air quality such as: ``How's the weather this hour?", the LLM correctly parses the user's intent. Instead of relying on its pre-trained knowledge, it autonomously selects and invokes the \textit{`get\_recent\_sensor\_data'} tool through the MCP. This tool fetches the latest measurements directly from the database. The LLM then receives this structured data as context and synthesizes it into a concise, human-readable summary for the user. This process effectively bridges the gap between raw sensor data and actionable user insights, successfully mitigating the risk of data hallucination.

\begin{figure}[tb]
    \centering
    \includegraphics[width= 0.8 \columnwidth]{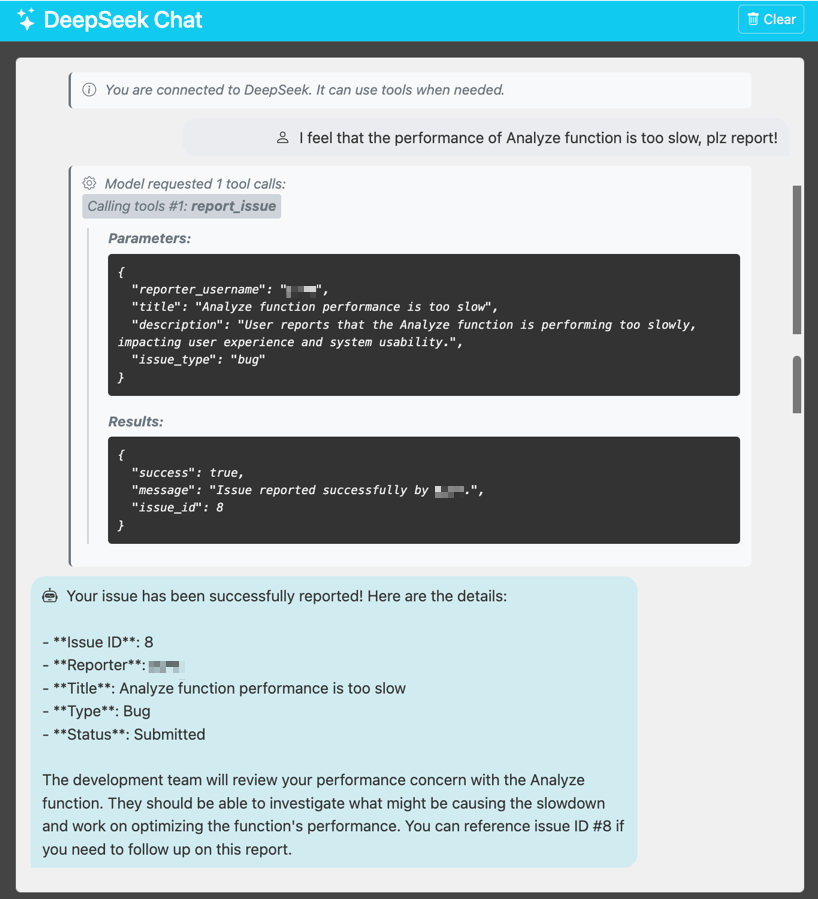}
    \caption{Demonstration of a system operation task. LLM interprets a user's issue report and calls the \textit{`report\_issue'} tool to create a ticket in the backend system.}
    \label{fig:issue_report}
\end{figure}

Beyond simple data retrieval, the AMI platform empowers the LLM to perform system-level operations. As shown in Figure~\ref{fig:issue_report}, a user can report a problem directly through the chat interface. The LLM is capable of understanding this request, extracting relevant information from the conversation, and calling the appropriate \textit{`report\_issue'} tool. The MCP server then executes this function, creating a new issue ticket (e.g., Issue \#8) in the system's backend. The final confirmation is relayed back to the user in natural language. This demonstrates how the LLM acts as an intelligent intermediary, streamlining user support and system management tasks.

\begin{figure}[tb]
    \centering
    \includegraphics[width=0.9\columnwidth]{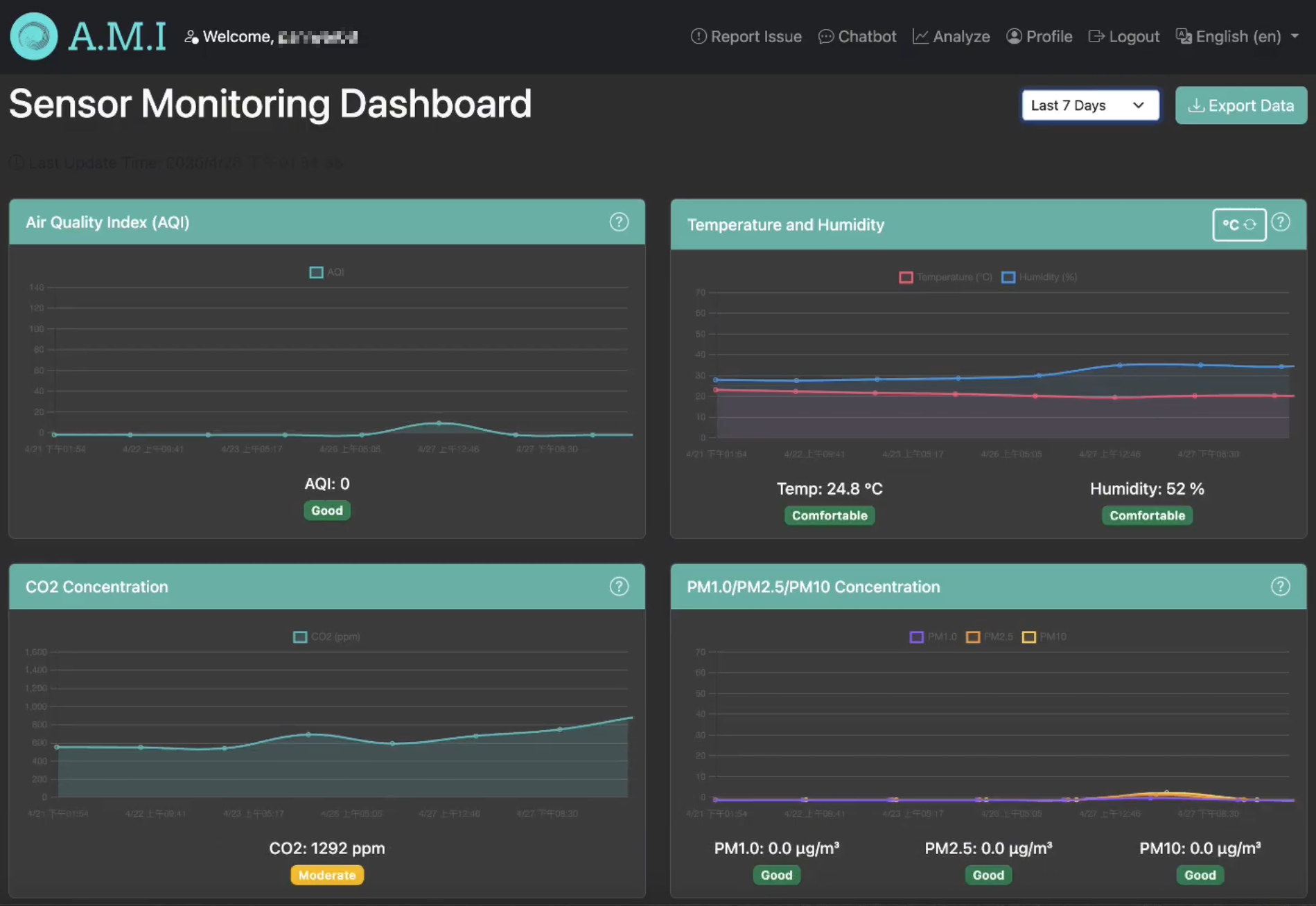}
    \caption{A snippet of AMI monitoring dashboard with data export feature}
    \label{fig:dashboard}
\end{figure}

Our system design incorporates a \textit{Defense-in-Depth} security architecture to handle more advanced interactions, such as modifying user data. When a user requests to update their profile, LLM calls the \textit{`update\_user\_profile'} tool. Even if the LLM were to be manipulated into trying to change another user's data, our backend implementation enforces a crucial security check: if forcibly overwrites the user identifier in the tool call to match the currently authenticated user. This ensures that the operation is always performed within the correct and authorized scope. The system then successfully updates the database then and provides a confirmation to the user. This dual mechanism, combining LLM tool-calling with strict backend validation, demonstrates a robust and secure approach to granting LLMs control over sensitive operations. As part of the overall system interface, Figure~\ref{fig:dashboard} highlights the developed user-friendly AMI monitoring dashboard, which complements these interactions by providing a visual overview and data export capability.

\subsection{Inter-rater Reliability Analysis}
We further performed a quantitative assessment of agreement among evaluators by computing inter-rater reliability (IRR). Inter-rater reliability measures the degree of consistency among different evaluators who assess the same set of responses \cite{mchugh2012interrater}. In this study, IRR is crucial because it ensures that the performance scores assigned to system outputs are not biased by individual subjectivity, but instead reflect a consistent and reproducible judgment across multiple raters. The evaluation was conducted based on four major criteria: Factual Accuracy, Completeness, No Hallucination, and Usefulness. Each criterion was rated on a five-point Likert scale, where a score of 5 indicated the highest quality labeled as ``Completely Accurate" and a score of 1 indicated the lowest quality labeled as ``Completely Inaccurate", with intermediate points representing varying degrees of accuracy and reliability. Six expert reviewers independently evaluated a total of 15 responses generated by the model, assigning scores for each criterion. This structured approach ensured that the assessment captured multiple dimensions of performance while providing a standardized framework for comparing judgments across raters. The aggregated results of the evaluation are presented in Table \ref{tab:evaluation_results}.

As shown in Table \ref{tab:evaluation_results}, we report both the average ratings and multiple IRR metrics to capture different dimensions of reviewer agreement. Average scores across the four criteria were consistently high, ranging from 4.74 for `Usefulness' to 4.84 for `No Hallucination', indicating strong overall quality. To complement these averages, we computed three agreement metrics. Weighted Kappa accounts for the ordinal nature of the five-point scale, treating near agreements (e.g., 4 vs. 5) as closer than distant ones. The Intraclass Correlation Coefficient (ICC 3,1) evaluates consistency when raters are considered as a fixed group. The Mean Absolute Difference (MAD) provides an intuitive measure of disagreement, where lower values indicate closer alignment. `No Hallucination' achieved the highest average rating of 4.84 and strongest agreement (Kappa 0.234, ICC 0.350, MAD 0.213). `Factual Accuracy' and `Completeness' similarly attained high scores of 4.78 and 4.82 respectively, with moderate agreement. `Usefulness', while slightly more variable with a MAD score of 0.413, still reflected general consensus despite its subjectivity. Taken together, these results highlight both the strong quality of outputs and the robustness of the evaluation. However, when averaged across all criteria, Kappa values (0.058–0.234) and ICC (0.037–0.350) suggest slight to fair agreement overall, a contrast that will be further examined in the discussion.

\begin{table}[tb]
\caption{Inter-rater reliability (IRR) metrics across evaluation criteria}
  \centering
  \setlength{\tabcolsep}{3pt}
  \begin{tabular}{lcccc}
    \toprule
     Criterion & Avg & Weighted Kappa & ICC (3,1) & MAD \\
    \midrule
    Factual Accuracy & 4.78 & 0.098 & 0.109 & 0.364 \\
    Completeness     & 4.82 & 0.077 & 0.044 & 0.284 \\
    No Hallucination & \textbf{4.84} & 0.234 & 0.350 & \textbf{0.213} \\
    Usefulness       & 4.74 & 0.058 & 0.037 & 0.413 \\
    \bottomrule
  \end{tabular}
  \label{tab:evaluation_results}
\end{table}

\section{Discussion}

The evaluation results provide strong evidence for the system's functional success, while also highlighting the nuanced challenges of evaluating generative AI systems. The primary finding is the co-occurrence of very high mean scores and relatively low inter-rater reliability (IRR) coefficients. While the average ratings across all four criteria were consistently high, the agreement measures such as Weighted Kappa and ICC were notably lower. This apparent contradiction is a well-known statistical artifact in reliability research, often described as the ``Kappa paradox" or ``ceiling effect" \cite{feinstein1990high}. When evaluators consistently give high ratings clustered at the upper end of the scale (scores of 4 or 5), the lack of variability reduces the sensitivity of reliability coefficients, making them appear lower even in the presence of substantive consensus. Several methodological factors also contribute to this outcome. The evaluation was based on only four criteria, which, while central to system performance, do not capture the full range of qualitative variation. With such limited criteria, evaluators may apply slightly different interpretations that reduce consistency. These issues combined highlight the difficulty of applying conventional IRR metrics in settings where system outputs are of uniformly high quality.

Despite these statistical nuances, the consistently high mean scores directly validate our research contributions. The near-perfect rating for Absence of Hallucination (4.84) demonstrates the success of our MCP-based architecture, which grounded the model’s outputs through tool calls for live data retrieval and mitigated the generation of fabricated information. Likewise, the strong scores for Completeness (4.82) and Usefulness (4.74) confirm the practical value of our overall system design, showing that experts found the outputs both comprehensive and actionable. Taken together, these findings reinforce that the system not only achieves functional accuracy but also addresses key challenges of hallucination and user relevance in real-world applications, while also motivating further refinement of evaluation frameworks to better capture consensus in high-performing systems.



\section{Conclusion and Future Work}
This work addresses the dual challenges of making complex IoT data accessible to non-expert users and mitigating the inherent risk of hallucination in LLMs. We developed the Air Monitoring Interface (AMI), a novel system that integrates an LLM with a real-time data back-end via the MCP server protocol. Expert evaluation showed high factual accuracy and usefulness, with the MCP-based architecture effectively grounding responses in live data and virtually eliminating hallucinations. The contributions of this work span four dimensions. First, unlike direct data injection methods limited by context windows, AMI leverages MCP to provide on-demand access to real-time sensor data. Second, unlike RAG pipelines that often underperform on structured time-series data, AMI enables the LLM to operate as an agent capable of invoking precise backend tools. Third, by adopting MCP as a standardized and decoupled integration layer, the system moves beyond ad hoc APIs, improving extensibility and maintainability. Finally, our \textit{defense-in-depth} design, combining prompt engineering with enforced code-level authentication, demonstrates a principled approach to deploying LLM-powered IoT systems securely and reliably.

Looking ahead, we acknowledge the limitations of our study, particularly the small cohort of expert evaluators. Future work will extend AMI by expanding the MCP toolset to support more complex analytical tasks, integrating additional LLMs for interoperability testing, and conducting large-scale usability studies with diverse participants. These efforts will further validate the system's practical impact and advance the broader vision of building accessible, trustworthy, and intelligent interfaces for real-time environmental monitoring.

\section*{Acknowledgment}
We would like to extend our sincere gratitude to Yen-Lin Chang, Fang-Lue Ju, Chung-Kun Wu, Collin Brey, Carson K. Lissome, Matthew Scott Maijala, and Taiwo Abe for all their invaluable help and contributions to this research.

\bibliographystyle{IEEEtran}
\bibliography{Ref}
\vspace{12pt}

\end{document}